\documentclass[floatfix,twocolumn,amsmath,prb]{revtex4}
\usepackage{color}
\usepackage{multirow}
\usepackage{graphicx}
\usepackage{amsmath}
\usepackage{amssymb}
\usepackage{bm}
\usepackage{dcolumn}
\usepackage{braket}

\begin{document}

\title{Engineering quantum anomalous Hall phases with orbital and spin degrees of freedom}

\author{Hongbin~Zhang}
\email[corresp.\ author: ]{h.zhang@fz-juelich.de}
\author{Frank Freimuth}
\author{Gustav Bihlmayer}
\author{Marjana Le\v{z}ai\'c}
\author{Stefan~Bl\"ugel}
\author{Yuriy~Mokrousov}
\email[corresp.\ author: ]{y.mokrousov@fz-juelich.de}
\affiliation{Peter Gr\"unberg Institut and Institute for Advanced Simulation, 
Forschungszentrum J\"ulich and JARA, D-52425 J\"ulich, Germany}

\date{today}

\begin{abstract} 
Combining tight-binding models and first principles calculations, we investigate the quantum anomalous Hall (QAH) effect
induced by intrinsic spin-orbit coupling (SOC) in buckled honeycomb lattice with $sp$ orbitals in an external exchange field. 
Detailed analysis reveals that nontrivial topological properties can arise utilizing 
not only spin but also orbital degrees of freedom in the strong SOC limit, when the bands acquire non-zero Chern numbers
upon undergoing the so-called {\it orbital purification}. As a prototype of a buckled honeycomb lattice with strong SOC we choose
the Bi(111) bilayer, analyzing its topological properties in detail. In particular, we show the emergence of several QAH phases
upon spin exchange of the Chern numbers as a function of SOC strength and magnitude of the exchange field. Interestingly,
we observe that in one of such phases, namely, in the {\it quantum spin Chern insulator} phase, the quantized charge and spin Hall 
conductivities co-exist. We consider the possibility of tuning the SOC strength in Bi bilayer via alloying with isoelectronic Sb, and
speculate that exotic properties could be expected in such an alloyed system owing to the competition of the topological 
properties of its constituents. Finally, we demonstrate that $3d$ dopants can be used to induce a sizeable exchange field
in Bi(111) bilayer, resulting in non-trivial Chern insulator properties. 

\end{abstract}

\maketitle

\section{introduction}

In recent years, dissipationless charge and spin transport phenomena have drawn very intensive 
attention.  After the topological nature of the anomalous Hall effect (AHE) in ferromagnets was discovered
and understood,\cite{Nagaosa:2010} the existence of the spin Hall\cite{Hirsch:1999, Kato:2004} 
and quantum spin Hall (QSH) effects\cite{Kane:2005,Konig:2007} was theoretically proposed and 
experimentally demonstrated. Nevertheless, the experimental observation of the so-called quantum anomalous 
Hall effect (QAHE),\cite{Haldane:1988} characterized by a quantization of the anomalous Hall conductivity, 
is still missing, despite various theoretical proposals\cite{Liu:2008,Yu:2010,Qiao:2010,Zhang:2011a,Zhang:2012a} 
and recent experimental advances in this direction.\cite{Chang:2011}

Numerous studies on the QSH effect have spawned a fascinating field focusing on the topologically
nontrivial states of matter, in particular, topological insulators (TIs).\cite{Hasan:2010, Qi:2011}  
Remarkable physical properties of TIs originate from the gapless surface states, guaranteed by 
a finite bulk gap and time-reversal symmetry. It is plausible to seek QAH effects by 
introducing time-reversal broken perturbations to TIs while keeping the topological
nontriviality,~e.g.~via a proximity effect.~\cite{Qi:2011}
Among all known TIs,~\cite{Yan:2012} two-dimensional (2D) systems are of particular interest,
such as graphene,~\cite{Kane:2005} HgTe quantum wells,~\cite{Konig:2007} AlSb/InAs/GaSb quantum wells,~\cite{Liu:2008a} 
silicene,~\cite{Liu:2011a} and Bi(111) bilayers (BLs).~\cite{Murakami:2006, Wada:2011}
In the past, a lot of work has been done based on the extended Kane-Mele model~\cite{Kane:2005}
in the context of graphene.\cite{Qiao:2010, Tse:2011, Chen:2011,  Qiao:2012, Ezawa:2012,Beugeling:2012}
In these studies, only the $\sigma$ band of the $p_z$ orbitals are considered, with key terms
to induce topological nontriviality being effective projected spin-orbit and Rashba spin-orbit interactions,
arising from second-order perturbative processes.~\cite{Min:2006} The strength of intrinsic atomic spin-orbit coupling of carbon 
is limited, though it can be enhanced in fine-tuned systems.~\cite{Weeks:2011}

In the quantized case of the AHE in two dimensions, the value of the anomalous Hall conductivity (AHC) is proportional to the so-called (first) Chern 
number, obtained as an integral of the Berry curvature of the occupied states in the Brillouin zone. This integer Chern
number, which characterizes the ``phase" complexity of the manifold of Bloch electrons as a whole, has its roots in solid 
state physics since the discovery of the quantum Hall effect  and its theoretical understanding by Thouless {\it et al.}
in 1982.~\cite{Thouless:1982} Shortly after that, Haldane suggested a spinless model to realize the QAHE without an external magnetic
field,~\cite{Haldane:1988} an effort, which conceptually led to the theoretical suggestion of graphene as a first topological insulator.~\cite{Kane:2005}
In the latter
case, the Chern numbers of spin-up and spin-down electrons are non-zero due to a microscopic realization of the Haldane's 
model, but they are opposite to each other. The generality of this counteracting nature of the Chern numbers in TIs lead eventually to the
concept of the spin Chern numbers, which can be used to characterize the TI phases in time-reversal broken situations.~\cite{Prodan:2011}
Ultimately, the realization of the QAHE is hinged on the presence of points in the electronic structure, at which 
otherwise separate spin bands interact with each other, and exchange corresponding spin Chern numbers, to give a global
non-vanishing Chern number in the system. The mechanism behind such a Chern number exchange is strongly dependent
on the details of symmetry, hybridization and microscopic structure of the spin-orbit interaction in the system, making
search for systems which could exhibit non-zero QAHE difficult, yet definitely rewarding since the physics involved
is possibly very rich. 


Among other materials, Bismuth (111) bilayer (BL) has been shown to host nontrivial topological properties due to the strong SOC of Bi.~\cite{Murakami:2006, Wada:2011,Zhang:2012} 
In our previous work, we used the concept of the spin Chern number to characterize the topological
phase transitions in Bi(111) bilayers with respect to the strength of an external exchange field.\cite{Zhang:2012}
Here, we extend this work further, and demonstrate the microscopic origin of the QAH phases found in Bi(111) bilayers
when utilizing both orbital and spin degrees of freedom. Based on a general tight-binding model for buckled honeycomb lattices
 with $sp$-orbitals, we illustrate different Haldane-inspired\cite{Haldane:1988} mechanisms which can lead to a nonzero Chern 
 number in each individual band upon employing the spin and orbital complexity (section II). We find that in addition to $p_z$ orbital based physics, in the strong
SOC limit, orbital angular momentum purification can also lead to nontrivial topological properties. We further clarify the ways
behind  achieving the  QAHE at the so-called spin-mixing points in the electronic structure. Finally, using first-principles
techniques, we show that various topological phases occur in a  realization of a buckled $sp$ honeycomb model $-$ the 
Bi(111) bilayer $-$ with respect to the SOC strength and magnitude of an external exchange field (section III). In particular we suggest 
the existence of a so-called quantum spin Chern insulator phase, which is characterized by the quantization of both the 
transverse charge and spin conductivities. We conclude by discussing how to realize the tuning of SOC strength by alloying Bi 
with Sb, and the effect of an exchange field by introducing magnetic substrates and/or dopants.

\section{Model analysis}

To illustrate how nonzero Chern numbers and nontrivial QAH phases can arise, we first
consider a generic tight-binding Hamiltonian of $sp$-orbitals on a  buckled honeycomb lattice
in $x$-$y$ plane:
\begin{equation}\label{Eq1}
H\ =\ \sum_{i,j}t_{i,j}c^\dagger_ic_j + \sum_ic_i^\dagger(\epsilon_i\Bbb{I}+Bs_z)c_i\ +H_{\text{SOC}} 
\end{equation}
where the first term presents the kinetic hopping with  $t_{i,j}$ ($i,j=s,p_x,p_y,p_z$) as the hopping parameters. 
The second term reflects an orbital-dependent on-site energy $\epsilon_i$ and the interaction with the Zeeman 
exchange field $B$ directed along the $z$-axis, with $\mathbb{I}$ ($s_z$) as the identity (Pauli) matrix. 
The third term in Hamiltonian~(\ref{Eq1}) is the on-site SOC Hamiltonian.
In order to facilitate the analysis, $s$, $p_z$, and \{$p_x,p_y$\} bands are artificially separated from each other by 
imposing a rigid shift of $\epsilon_i$. To identify different origins of the QAHE, the spin-orbit interaction is
further decomposed into spin-conserving and spin-flip parts:
\begin{equation}\label{Eq2}
H_{\text{SOC}}\ =\ \xi\mathbf{l}\cdot\mathbf{s}=\xi l_zs_z+\xi(l^+s^-+l^-s^+)/2
\end{equation}
where $\mathbf{l}$ ($\mathbf{s}$) is the orbital (spin) angular momentum operator, and $\xi$ is the 
electron-shell averaged atomic SOC strength. Since in this work we choose the direction of the spin-polarization 
to be aligned along $z$ direction, the spin conserving part of the SOC Hamiltonian, $\xi l_zs_z$,
couples \{$p_x,p_y$\} orbitals, while the spin flip part of Eq.~\ref{Eq2}, $\xi(l^+s^-+l^-s^+)/2$, couples $p_z$ and 
\{$p_x,p_y$\} orbitals via a flip of spin and a $\pm 1$ change in the orbital quantum number. Further, 
we take the Slater-Koster tight-binding parameters of Bi, provided in Ref.~[\onlinecite{Liu:1995}] for the values of
hopping integrals of the model, $t_{i,j}$.

Before proceeding further with the analysis of Eq.~(\ref{Eq1}), we would like to remind the reader of the Haldane 
model,\cite{Haldane:1988} which is a generic model exhibiting an emergence of the QAH effect on a honeycomb 
lattice  without an external magnetic field. The key mechanism in this model is the complex hopping 
between the next nearest neighbor (NNN) orbitals, which leads to a sublattice-dependent ``magnetic field" with 
zero flux through the overall hexagonal plaquette. In our work, it is the on-site atomic SOC that induces such 
complex hopping via different processes as discussed below.

\subsection{$p_z$ orbitals}

\begin{figure*}[ht]
\includegraphics[width=17cm]{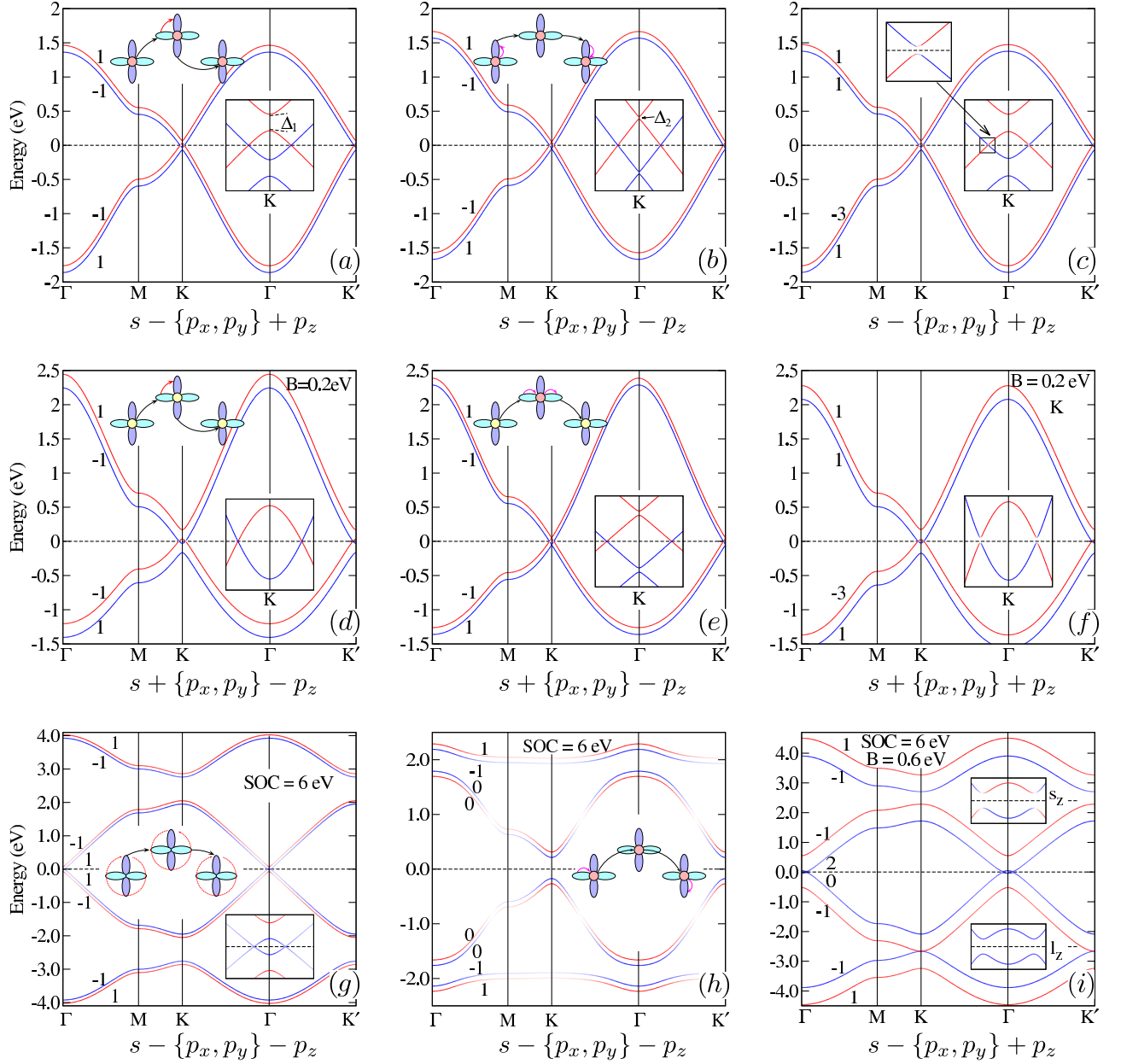}
\caption{Topological analysis of $p_z$ ((a)-(c)), $s$ ((d)-(f)), and \{$p_x,p_y$\} ((g)-(i)) 
bands in a buckled honeycomb bilayer.
The electronic structure is obtained using the tight-binding parameters for bulk Bi as listed in 
Ref.~[\onlinecite{Liu:1995}], with an additional artificial rigid shift of
on-site energies yielding separation of $s$, $p_z$, \{$p_x,p_y$\} states in energy. 
Only nearest neighbor (NN) hoppings are considered, with V$_{pp\pi}=0$ 
(this does not affect our conclusions, see the main text). 
Without specification, the spin-orbit coupling parameter $\xi=4.5$ eV and exchange field $B=0.1$ eV are used.
Left column (a,d,g) (middle column (b,e,h)) corresponds to the 
case with only spin-conserving SOC (spin-flip SOC) included, while the full SOC is considered in the right column (c,f,i). 
Red (blue) in (a-f) and upper inset of (i) stand for the spin-down 
(spin-up) states, while in (g-i) and lower inset of (i) for the states with orbital magnetic number $m=+1$ ($m=-1$).
Dashed horizontal lines indicate the Fermi energy level adjusted such that relevant bands (for instance, $p_z$ in (a-c)) 
are half-filled. 
Numbers denote the Chern number of each individual band.
Insets in (a-g) ((i)) display the electronic structure at the Dirac critical 
point at $\pm 0.01$ eV ($\pm 0.06$ eV) with respect to the Fermi energy, 
and sketches illustrate different channels for complex nearest neighbor hopping 
(red (yellow) circles denote $p_z$ ($s$) orbitals, while $p_{x,y}$ orbitals are indicated by blue ellipsoids; 
black (red) arrows depicts $t_\text{NN}$ hoppings (SOC hybridization), respectively.).
Text below each panel denotes the hybridization of \{$p_x,p_y$\} orbitals to $s$ and $p_z$ orbitals 
by kinetic hoppings, where $+$ ($-$) indicates (no) hybridization.
For instance, ``$s-\{p_x,p_y\}+p_z$" in (a),(i) means that there is no 
hybridization between $s$ and \{$p_x,p_y$\}, while \{$p_x,p_y$\} and $p_z$ are coupled.
}
\label{fig:model}
\end{figure*}

First, we consider the case of the $p_z$ bands, particularly relevant in graphene physics, well-separated from 
\{$p_x,p_y$\} and $s$ states. 
Generally, there are two ways to induce nontrivial Chern numbers in 
the $p_z$ bands via generating an effective next nearest neighbor (NNN) complex hopping due to intrinsic atomic SOC.
First, on a buckled honeycomb lattice, $p_z$
orbitals can hybridize directly with the \{$p_x,p_y$\} orbitals, and complex hoppings can be induced via 
the spin-conserving 
part of SOC which acts between $p_x$ and $p_y$ states. As illustrated in Fig.~\ref{fig:model}(a), in this mechanism
the corresponding virtual transitions read:\cite{Liu:2011b}
\begin{equation}
\mid p_z^\text{A}\uparrow\rangle\overset{t_{\text{NN}}}{\rightarrow}\mid p_{x,y}^\text{B}\uparrow\rangle\overset{\xi l_zs_z}{\rightarrow}\mid p_{x,y}^\text{B}\uparrow\rangle
\overset{t_\text{NN}}{\rightarrow}\mid p_z^\text{A}\uparrow\rangle
\end{equation}
where $V$ indicates the direct hopping between $p_z$ and $p_{x,y}$ orbitals on the neighboring sites, while 
superscripts A and B
denote the nearest neighbor atomic sites in sublattice A  and B. 

For the resulting NNN hopping we obtain $t_{\text{NNN}}\sim\xi$.
Importantly, the effective hoppings are opposite in sign for $p_z$ orbitals of a given spin on A and B sublattice, hence giving rise to a 
finite gap $\Delta_1\sim2\xi$ at the K (K$^\prime$) high-symmetry points in the Brillouin zone, as 
shown in Fig.~\ref{fig:model}(a), in which a small magnetic field was applied in order to separate the spin-up and spin-down
bands in energy. The gap is topologically nontrivial, in direct accordance to Haldane's arguments.\cite{Haldane:1988} The two 
spin-channels here are independent, since only the spin-conserving part of SOC is involved, although the complex NNN hoppings 
for the spin-down and spin-up electrons are complex conjugate of each other.\cite{Liu:2011b} This means that the Chern numbers of the spin-up bands are opposite to the Chern numbers of
the spin-down bands, as shown in Fig.~\ref{fig:model}(a), which suppresses the QAHE of occupied electrons.

Secondly, on-site spin-flip SOC can give rise to complex next neighbor hopping too, even if there is no direct 
hybridization between $p_z$ and \{$p_x,p_y$\} orbitals, Fig.~\ref{fig:model}(b).
In this case, the corresponding virtual transitions are:~\cite{Min:2006, Konschuh:2010, Liu:2011b}
\begin{equation}
\mid p_z^\text{A}\uparrow\rangle\overset{\xi_\text{flip}}{\rightarrow}\mid p_{x,y}^\text{A}\downarrow\rangle\overset{t_{\text{NN}}}{\rightarrow}\mid p_{x,y}^\text{B}\downarrow\rangle\overset{t_\text{NN}}{\rightarrow}\mid p_{x,y}^\text{A}\downarrow\rangle
\overset{\xi_\text{flip}}{\rightarrow}\mid p_z^\text{A}\uparrow\rangle
\end{equation}
where $\xi_\text{flip}=\xi(l^+s^-+l^-s^+)/2$, and $V$ stands for the direct hybridization between $p_{x,y}$ orbitals on neighboring A and B sites.
In analogy to the case with spin-conserving SOC considered previously, the effective hoppings within A and B sublattices are of opposite sign and the resulting gap is also topologically nontrivial.
In this case  $t_{\text{NNN}}\sim\xi^2$, and the corresponding gap $\Delta_2$ 
at K (K$^\prime$)  (Fig.~\ref{fig:model}(b)) scales quadratically with
respect to the SOC strength.
If there is no hybridization between the $p_z$ states of the opposite spin as what occurs in the 
2D planar honeycomb lattice, this second situation without considering an external exchange field actually presents the topological insulator state as found in graphene,\cite{Kane:2005}  consisting of two independent spin copies of the 
Haldane model. Thus, the resulted QAH conductivity for the half-filled situation is again zero, although the $p_z$ bands of opposite spin are meeting at the Fermi energy.

Note that, while the arising non-zero Chern numbers of the bands in all situations discussed in Fig.~1 could lead to
a QAH effect upon inducing a very large (larger than the bandwidth) exchange splitting in the system, we do not discuss 
such ``trivial" possibility here.  
Thus, to induce nontrivial QAHE in the $p_z$ bands, the spin-up and spin-down $p_z$ bands have to be coupled, inducing exchange of the Chern numbers.
This can be achieved by allowing for the hybridization between $p_z$ and \{$p_x,p_y$\} orbitals and upon
taking the complete SOC into consideration.
When two entangled bands of different spin character from conduction and valence bands overlap with each other,
spin-mixing occurs and the Chern number exchange takes place when the gap is opened,
Fig.~\ref{fig:model}(c). For example, for $p_z$ orbitals, a possible transition process is:
\begin{equation}
\mid p_z^\text{A}\uparrow\rangle\overset{\xi_\text{flip}}{\rightarrow}\mid p_{x,y}^\text{A}\downarrow\rangle
\overset{\xi_{l_zs_z}}{\rightarrow}\mid p_z^\text{A}\downarrow\rangle.
\end{equation}
The magnitude of the gap, opened between up and down $p_z$-bands, as shown in Fig.~\ref{fig:model}(c), 
is proportional to $\xi^2$ according to our calculations.

The variation of Chern numbers upon band touching is determined by 
the Berry indices at degenerate points in a generalized parameter space ($\mathbf{k},\eta$), 
which is defined as:~\cite{Bellissard:1995,Lee:1998}
\begin{equation}
\text{Indx}_{\text{Berry}}=\frac{1}{2\pi}\int_{\mathbf{S}^2} ds\,\mathbf{\Omega}\cdot\mathbf{n},
\end{equation}
with $\mathbf{S}^2$ as a two dimensional surface enclosing the degeneracy point, $\mathbf{n}$ is the 
surface normal vector pointing outwards, $\Omega$ is the Berry curvature.~\cite{Lee:1998}
In our case, $\eta$ can be the strength of the exchange fields or SOC.
Since $\mathbf{S}^2$ can be chosen arbitrarily small, the magnitude of $\text{Indx}_{\text{Berry}}$ can be 
obtained by examining the band dispersion at the degeneracy point.~\cite{Bellissard:1995,Lee:1998} 
For instance, if the dispersion of the crossing bands is linear (quadratic), the variation of the Chern 
number is one (two). 
This saves us from evaluating the Chern number exchange explicitly, although the sign of $\text{Indx}_{\text{Berry}}$ cannot be determined
from this simple argument.
In the case of Fig.~\ref{fig:model}(c), the Chern number of the upper valence band is 
reduced by 2, while the Chern number of the lower conduction band is increased by 2, as compared to the situation in
Figs.~\ref{fig:model}(a) and (b), when the hybridization between the $p_z$ bands of opposite spins is introduced. The reason for this is that, although the dispersion of the bands at the point of degeneracy is obviously linear, Fig.~\ref{fig:model}(c), there are two such points in the Brillouin zone (K and K').

In previous studies of the QAHE in the context of graphene,\cite{Tse:2011,Chen:2011,Qiao:2012,Beugeling:2012,Ezawa:2012}
an extended Kane-Mele model~\cite{Kane:2005} was employed with an additional Rashba term.
The effective Rashba spin-orbit interaction originates from the combination of intrinsic SOC and potential gradient perpendicular
to the honeycomb lattice plane which breaks the inversion symmetry, and can be created by~e.g.~applying a finite electric field perpendicular
to the honeycomb plane\cite{Min:2006} or imposing a finite curvature of the honeycomb sheet.\cite{Hernando:2006} 
In this case, when the time-reversal symmetry is broken via,~e.g., non-zero magnetization due to adatoms, the QAHE can be induced.\cite{Qiao:2010,Zhang:2012a} Nevertheless, the effective complex hoppings due to
the Rashba spin-orbit arise between $p_z$ orbitals of opposite spin on the nearest neighbor sites $-$ a situation,
topologically distinct from the complex NNN hoppings considered in this work.

\subsection{$s$ orbitals}

The second-order perturbation processes discussed above can also take place if the $s$ bands are taken instead of 
$p_z$ states.  It is achieved by hybridization between $s$ and $p$ orbitals, Fig.~\ref{fig:model}(d)-(f).
For instance, the effective complex hoppings between $s$ orbitals can be induced by spin-conserving 
SOC via hybridizing with \{$p_x,p_y$\} orbitals (Fig.~\ref{fig:model}(d)) or spin-flip SOC (Fig.~\ref{fig:model}(e)).
Comparing to the $p_z$ orbitals, the only difference is that for the case with spin-flip SOC (Fig.~\ref{fig:model}(e)), 
the corresponding virtual transitions are
\begin{equation}
\mid s^\text{A}\uparrow\rangle\overset{t_\text{NN}}{\rightarrow}\mid p_{x,y}^\text{B}\uparrow\rangle\overset{\xi_\text{flip}}{\rightarrow}\mid p_{z}^\text{B}\downarrow\rangle\overset{\xi_\text{flip}}{\rightarrow}\mid p_{x,y}^\text{B}\uparrow\rangle
\overset{t_\text{NN}}{\rightarrow}\mid s^\text{A}\uparrow\rangle.
\end{equation}
Note that the role of $\mid p_{x,y}^\text{B}\rangle$ and $\mid p_z^\text{B}\rangle$ in this case can be exchanged,~i.e., electrons can hop from $\mid s^\text{A}\rangle$ to $\mid p_z^\text{B}\rangle$ and then couple with $\mid p_{x,y}^\text{B}\rangle$.
Furthermore, similarly to the cases for $p_z$ bands (Fig.~\ref{fig:model}(a-c)), the spin-up and spin-down $s$ bands 
are decoupled with Chern numbers of opposite
sign (Fig.~\ref{fig:model}(d) and (e)), and a nontrivial QAH effect for the half-filled case takes place only when
the two spin-channels are entangled,~i.e., due to hybridization with $p_z$ orbitals (Fig.~\ref{fig:model}(f)).

\subsection{\{$p_x,p_y$\} orbitals: orbital purification}

More interestingly, nontrivial topological properties arise in the \{$p_x,p_y$\} bands as well, Fig.~\ref{fig:model}(g-i).
If only the spin-conserving SOC is considered, every \{$p_x,p_y$\} band acquires a nonzero Chern
number when the magnitude of SOC is larger than a critical value $\xi_{op}$ 
(Fig.~\ref{fig:model}(g)).
This is different from the cases considered for $p_z$ and $s$ bands, where the strength of SOC determines only the
size of the gap, while it is irrelevant for the topological properties of  individual bands.
Our analysis reveals that the origin of the nontrivial Chern numbers can be attributed to the so-called {\it orbital purification}, caused by strong SOC (Fig.~\ref{fig:model}(g)). For instance, for the middle four bands in Fig.~\ref{fig:model}(g), the Chern 
numbers are nonzero only when SOC is larger than $\xi_{op}=5.4$ eV, when the value of the orbital angular
momentum $L_z$ in each of the bands becomes consistently either strictly positive or negative at each $k$ 
point (see the coloring
of the bands in Fig.~\ref{fig:model}(g)). This leads to the predominantly $m=\pm 1$ ($p_x\pm ip_y$) character of each
band, which constitutes the essence of orbital purification. 
The complex NNN hoppings in this case are induced via the following mechanism:
\begin{equation}
\mid p_x^\text{A}\pm ip_y^\text{A}\rangle \overset{V}{\rightarrow} \mid p_x^\text{B}\mp ip_y^\text{B}\rangle \overset{V}{\rightarrow} \mid p_x^\text{A}\pm ip_y^\text{A}\rangle 
\end{equation}
due to virtual transitions between the bands of different $m$ on different sublattices, mediated by kinetic hopping. 
Since only spin-conserving SOC is considered here, the spin-up and spin-down channels are not entangled, if 
no direct hybridization of \{$p_x,p_y$\} and other orbitals is present, which corresponds to the situation discussed 
by Wu\cite{Wu:2008, Zhang:2011} for spinless honeycomb optical lattices. 

On the other hand, for the lowest (top most) two bands, the orbital angular momentum purification occurs for 
an even  smaller SOC strength of about $0.3$ eV, when only spin-conserving SOC is considered.
There exists also another possibility for these bands to acquire a complex
NNN hoppings by spin-flip SOC as shown in Fig.~\ref{fig:model}(h).
In this case, the Chern numbers for the middle four bands are always zero
even if the spin-flip SOC strength is increased to very large values, while for the lowest 
(top most) two bands, the Chern numbers are quantized with nonzero value.
To obtain a nontrivial QAHE, entanglement between two spin channels is again essential, where 
Chern number exchange occurs via spin-mixing as observed for the $s$ and $p_z$ cases above.
For \{$p_x,p_y$\} orbitals, such entanglement can be achieved by hybridization with $p_z$-orbitals,
with corresponding Chern number exchange of 1 due to a linear band dispersion at the points of degeneracy.
Following previous arguments, 
this leads to a non-zero QAH conductivity with $\mathcal{C}=-1$ at half-filling, Fig.~\ref{fig:model}(i).

\begin{figure}[ht]
\includegraphics[width=6cm, angle=270]{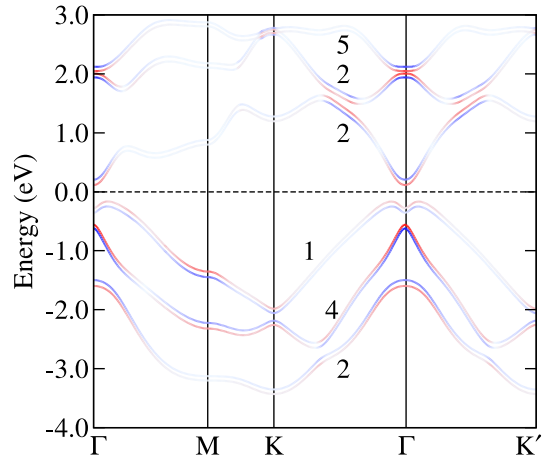}
\caption{Orbital angular momentum $L_z$ purification in Bi(111) bilayer from first principles calculations.
Red (blue) stands for the expectation value of the $L_z$ operator. A small artificial exchange field of $10$ meV 
was applied perturbatively to split originally degenerate
bands. In each pair of resulting nearly degenerate bands the expectation value of $L_z$, $s_z$ and the 
Chern numbers are opposite  to each other. Integers next to the bands stand for the absolute value of the 
Chern numbers.
\label{fig:orbital}
}
\end{figure}

The orbital purification is determined by the competition between the strength of the on-site SOC and the band width
due to kinetic hoppings. In the model analysis above, we neglected the V$_{pp\pi}$ hoppings
in Bi, with resulting $\xi_{op}$ of about $5.4$ eV. This critical value is significantly reduced if V$_{pp\pi}$ hoppings are 
taken into account, which reduces the band width of the $p$ bands. In the latter case, for Bi, $\xi_{op}$ becomes $2.4$ eV if all 
relevant hopping integrals for the bilayers, as listed in Ref.~[\onlinecite{Liu:1995}], are included in our tight-binding 
calculations.
Our first principles calculations of the Bi (Sb) bilayer, on the other hand, show that the magnitude of the atomic SOC for Bi, $\xi_{\rm Bi}$ 
($\xi_{\rm Sb}$ for Sb), is about $2.8$ eV ($0.9$ eV).
Thus, Bi bilayer resides in the strong SOC limit, where the orbital purification takes place, orbitally polarized $p_x\pm ip_y$ 
bands are formed and exhibit non-zero Chern numbers. To verify this explicitly, we evaluate the expectation value of 
the orbital momentum operator $L_z$ in Bi bands.
As shown in Fig.~\ref{fig:orbital}, the first and the third valence bands of Bi(111) bilayer are indeed orbitally purified 
in the entire Brillouin zone.
The corresponding Chern numbers are $\pm 2$ and $\pm 1$, respectively.
The orbital purification in these bands occurs until the SOC strength of Bi atoms is scaled down to 70\% of its atomic 
value, that is, approximately 2~eV, when the Bi bilayer looses its topological insulator properties, see Fig.~3(a). In this sense, 
we can call the Bi bilayer an {\it orbital topological insulator}, as opposed to the case of graphene, in which all the 
topological properties are due to $p_z$ states.

A comment on achieving nontrivial topological phases in optical lattices is in order, given that the idea of orbital 
angular moment purification was first proposed in the context of ultracold atoms.\cite{Wu:2008}
Quite recently, both Abelian~\cite{Lin:2009} and non-Abelian~\cite{Lin:2011,Wang:2012,Cheuk:2012} gauge fields 
(see Ref.~[\onlinecite{Dalibard:2011}] for a recent review) have been experimentally realized for trapped
neutral atoms. Since ultracold atoms are spinless particles, in this context the relevance of the models discussed here 
is limited to the cases shown in Fig.~\ref{fig:model}(a),(d),(g) when neglecting the spin degree of freedom.
Moreover, the hybridization between different orbitals (as in Fig.~\ref{fig:model}(a) and (d)) is not necessary, since 
in optical lattices complex hoppings are carried by dressed states. Experimentally, ultracold atoms in higher orbitals 
such as $p$ bands have been recently achieved (see e.g. Ref.~[\onlinecite{Wirth:2010}]), which makes them a 
promising candidate for realizing various topological properties. 


To summarize, we demonstrated that both orbital and spin degrees of freedom can be utilized to induce nonzero 
Chern numbers in an individual band, while nontrivial QAH effect is due to the entanglement of two spin channels 
accompanied by spin exchange of the Chern number. 
In real materials, all bands are in general coupled by hybridization and, hence, Chern number exchange is expected 
at every non-accidental band crossing. 
In this sense, graphene is a peculiar material,~\cite{Novoselov:2005}~since~the well-separated in energy $\sigma$ and $\pi$ bands are 
not coupled by kinetic hopping due to graphene's planar structure. Sublattice staggering
as found in silicene\cite{Liu:2011a} is too small to induce significant variation of the electronic structure, thus the 
simplified 4-band model considering only the $\pi$ bands is good enough to account for the topological phase transitions
in graphene-related systems.\cite{Tse:2011, Chen:2011, Qiao:2012, Ezawa:2012} However, for Bi(111) bilayer and its derivatives,
 all three $p$-orbitals reside in the same energy scale, the hybridization between them is strongly enhanced due to 
buckling, and the strength of SOC is orders of magnitude larger as compared to that of carbon or silicon atoms. 
Therefore, rich physics with competing orbital and spin degrees of freedom is expected in the latter case. 
In the next section, we will demonstrate that non-trivial QAH phases can be achieved in Bi(111) BL by applying an exchange 
field, where the variation of Chern numbers can be explained using the Chern number exchange scheme illustrated above.

\begin{figure*}[ht]
\includegraphics[width=17.95cm]{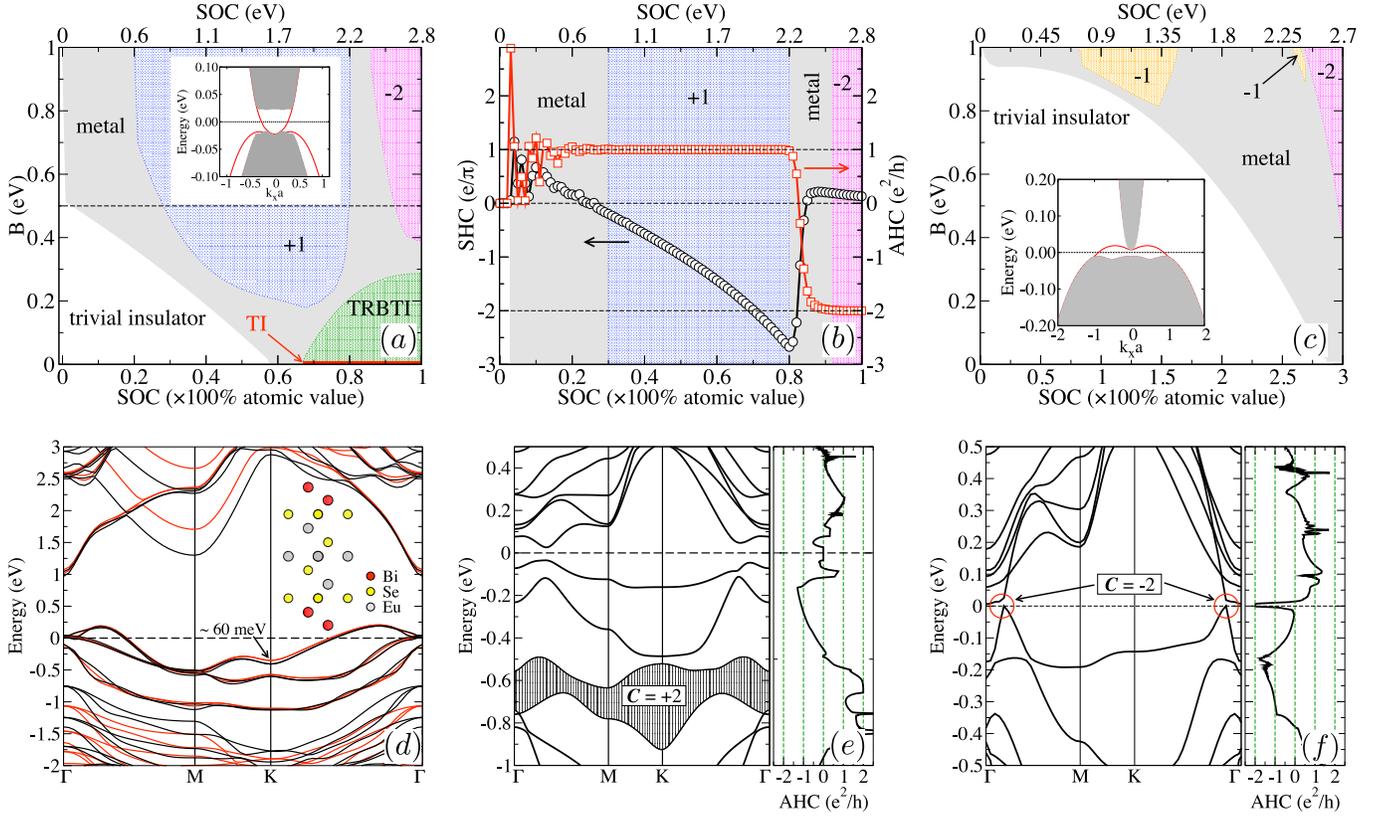}
\caption{Quantum anomalous Hall phases in Bi/Sb(111) bilayers, and electronic structure of the Bi(111) bilayer on magnetic 
substrates and with magnetic dopants.
(a) ((c)) displays the phase diagram of 
the Bi(111) (Sb(111)) bilayer with respect to the strength of atomic SOC and magnitude of exchange field $B$. Numbers denote the Chern
number in the QAH phase, ``TI" stands for the TI phase, while ``TRBTI" stands for the time-reversal broken TI phase. The horizontal dashed line 
in (a) indicates the case with $B=0.5$~eV, for which the AHC and spin Hall conductivity (SHC)
are shown in (b) with respect to the strength of SOC. Inset in (a) displays the dispersion of the edge states (red lines) and the projected bulk states (gray shaded region) in
Bi(111) BL ribbon with $B=1$ eV and  70\% of the Bi atomic SOC strength in the  $\mathcal{C}=+1$ phase, while that in (c) displays the dispersion of the edge states in Sb(111) BL ribbon with $B=1$ eV and 100\% of the atomic SOC of Sb in $\mathcal{C}=-1$ phase.
(d) displays the band structure of the Bi(111) BL on top of EuSe(111) terminated with Se atoms.
Black (red) lines denotes the majority (minority) bands. The exchange splitting at K point for the first two pairs of the valence bands
is about $60$ meV. The dashed horizontal line indicates $E_F$, and the inset displays the positions of the atoms.
Left panel of (e) shows the electronic structures of Bi(111) BL in $2\times2$ superstructure with one Fe atom located at the hollow site in between the two Bi layers, whereas AHC conductivity
is shown in the right panel. The horizontal dashed line in (d) indicates the position of the Fermi level which is not in the gap due to the fact that Se assimilate electrons. 
Shaded region in (e) marks the gap with the Chern number $\mathcal{C} = +2$. (f) is analogous to (e), but
with the SOC strength of Bi scaled down to 20\% of its atomic value.
\label{fig:fig2}
}
\end{figure*}

\section{Topological states of $\rm{Bi}$ and $\rm{Sb}$ bilayers}
In the previous section we illustrated within a tight-binding model how non-zero Chern numbers can arise due to intrinsic 
SOC and how to induce the QAH effect via applying an exchange
field. One problem remains unsolved, however, that is the problem of characterizing various topological phases consistently.
For instance, TIs are characterized by the $\mathbb{Z}_2$ index~\cite{Kane:2005a} and QAH insulators by the (first) Chern number, and 
it is still not completely clear how to properly characterize the 2D insulating topological phases which do not fit into the these two classes,
e.g.~insulators with broken time-reversal symmetry which originate from TIs and have a zero Chern number. Another issue is 
how such ``intermediate" phases can be probed and distinguished experimentally from topological and Chern insulators.
To this end,  we use the spin Chern number first introduced
by Sheng {\it et al.}~\cite{Sheng:2006} and later generalized to the cases with spin-flip SOC by Prodan.~\cite{Prodan:2011}  
Compared to the more universal approach of Ref.~[\onlinecite{Soluyanov:2012}],
the spin Chern number can be easily constructed. It is applicable to the situation without time-reversal symmetry,
and robust when the spectra of $P s_z P$ are gapped in the BZ, where $P$ is the projection operator onto 
the occupied states.
In the following, we use $\mathcal{C}_+$ ($\mathcal{C}_-$) to 
denote the ``Chern" number of the spin ``up" (``down") projected occupied bands, and the spin Chern number is defined as
$\mathcal{C}_\text{s}=\frac{1}{2}(\mathcal{C}_+-\mathcal{C}_-)$. For more details see Ref.~[\onlinecite{Zhang:2012}].

In this section we tackle the question of finding and characterizing possible topological phases of Bi and Sb bilayers from first principles,
as the corresponding Hamiltonian of these systems is altered by scaling the SOC matrix elements and applying an external exchange
field. Our {\it ab initio} calculations were performed using the full-potential linearized augmented plane 
wave method as implemented in the J\"ulich density functional theory code {\tt FLEUR}.\cite{fleur} 
The Wannier functions technique was used on top of self-consistent
first principles calculations to derive an accurate tight-binding Hamiltonian of the system.\cite{Souza:2002, Freimuth:2008, wannier90}
The relaxed bulk in-plane lattice constant and the distance between the two layers  for Sb(111) constitute 4.30~\AA~and 1.55~\AA,
respectively, while for Bi(111) bilayer the corresponding values are 4.52~\AA~and 1.67~\AA. 
The details of calculating the anomalous (spin) Hall conductivities, are described in Ref.~[\onlinecite{Zhang:2012}].
Moreover, we find that the spectra of $s_z$ projected onto the occupied
states are always globally gapped for the cases considered in the following, despite the fact that the strength of SOC in Bi, $\xi_{\rm Bi}$, is very 
strong ($\approx 2.8$ eV), leading to well-defined spin Chern numbers.
A uniform exchange field was applied on top of the first principles electronic structure, as described in Ref.~[\onlinecite{Zhang:2012}].
The spin-orbit strength $\xi$ in our calculations was scaled by hand via multiplying all atomic SOC matrix elements with a uniform 
scaling parameter during the
self-consistency cycle. In order to clarify the effect of a more realistic exchange field, we have performed additional calculations of a
Bi(111) bilayer on ferromagnetic substrates (europium chalcogenides) and in the presence of magnetic dopants in the system, as 
described later. 

\subsection{Phase diagram of Bi(111) bilayer}

The calculated phase diagram of the Bi(111) BL with respect to the strength of atomic SOC, $\xi$,  and the magnitude 
of the exchange field, $B$, is shown in Fig.~\ref{fig:fig2}(a). The emerging distinct topological phases can be characterized 
by the value of the AHC (which equals the Chern number times $e^2/h$ in the insulating regime) and the spin Chern number. 
The topological insulating phases are separated by a metallic phase, with topological phase transitions occurring during closing the
bulk band gap and reopening it again as $B$ and the  SOC strength are varied. 
As confirmed by the calculation of the spin Chern number, when the time-reversal symmetry is not broken ($B=0$ eV), 
the Bi(111) BL is a trivial insulator ($\mathcal{C}_\text{s}=0$, $\mathcal{C}=0$) for the SOC strength $\leq$ 56\% of 
its atomic Bi value, while it turns into a TI ($\mathcal{C}_\text{s}=-1$, $\mathcal{C}=0$) when $\xi$ is larger than 
67\% of $\xi_{\rm Bi}$.~\cite{Wada:2011}
In the TI phase, $\mathcal{C}_{+}=-\mathcal{C}_-=-1$, resulting in a spin Chern number of $-1$.

Breaking the time-reversal symmetry by applying an exchange field induces topological phase transitions;
If the magnitude of the induced exchange splitting is large enough 
to make the two bands, which were originally below and above the gap, to overlap, according to the mechanism 
depicted in Fig.~\ref{fig:model}(c),(f),(i). Before that, the original insulating phase is not destroyed, and to a certain extent
preserves its topological properties in the presence of a small exchange field. 
For example, for $0.2\,\xi_{\rm Bi}<\xi<0.56\,\xi_{\rm Bi}$ the trivial insulator phase
at $B=0$ retains as the $B$ is increased until the system enters the metallic phase in which the originally separated bands 
directly overlap.
On the other hand, for Bi BL with $\xi>0.67\,\xi_{\rm Bi}$, the original topologically nontrivial insulating gap is finite for finite values of $B$, 
with $\mathcal{C}_+=-\mathcal{C}_-=-1$ until entering the metallic region. The resulting spin Chern number is in this case $\mathcal{C}_\text{s}=-1$, 
which manifests the occurrence of the time-reversal broken TI (TRBTI) phase, observed also for graphene 
in~[\onlinecite{Yang:2011}], and discussed at length in~e.g.~[\onlinecite{Zhang:2012}].

The transition into the QAH phase upon increasing the $B$ always takes place via an intermediate metallic phase, Fig.~\ref{fig:fig2}(a). 
In our previous work we have shown the appearance of a QAH phase with $\mathcal{C}=-2$ for Bi BL with full atomic SOC and 
$B\geq 0.42$ eV.\cite{Zhang:2012} As we can see from Fig.~\ref{fig:fig2}(a), in a Bi BL with intermediate 
SOC strength, applying strong exchange field can also induce QAH phases which are derived from either a metallic, trivial insulator, 
or topological insulating phases at $B=0$. For instance, in a Bi BL with SOC strength scaled to one half of the atomic value, 
a QAH phase with $\mathcal{C}=+1$ emerges for $B\geq 0.25$ eV.
It is an exotic phase, in which $\mathcal{C}_-=+1$ while $\mathcal{C}_+ = 0$, as
compared to the QAH phase with $\mathcal{C} =-2$, in which $\mathcal{C}_- =\mathcal{C}_+ = -1$. 
In the former case, the spin Chern number $\mathcal{C}_s = \frac{1}{2}(\mathcal{C}_+ - \mathcal{C}_-)$
is not an integer, and there exist only one chiral edge state localized at each edge of a 
one-dimensional Bi(111) ribbon, as shown in the inset of Fig.~\ref{fig:fig2}(a). Our calculations show that these edge states are 
spin-polarized in the same direction, leading to the quantized value of the AHC (in units of $+e^2/h$)
and also finite and large values of the SHC, as demonstrated explicitly by calculations for the case of $B=0.5$ eV in 
Fig.~\ref{fig:fig2}(b). 

\begin{figure*}[ht]
\includegraphics[width=17cm]{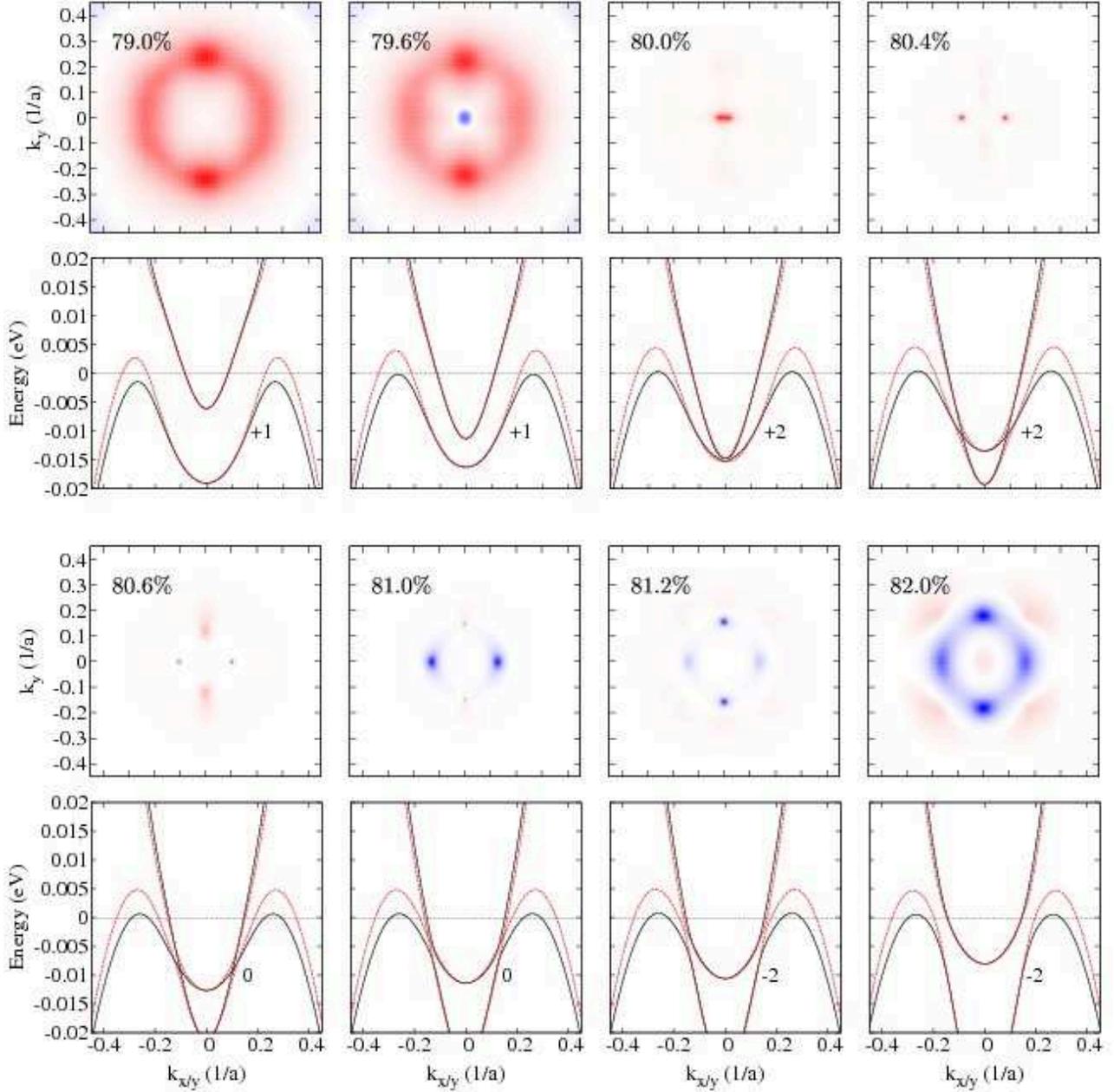}
\caption{Variation of the distribution of the Berry curvature (first and third row) and the band structure (second and fourth row) at the
$\Gamma$-point during a  phase transition from $\mathcal{C}=+1$ to
$\mathcal{C}=-2$ in Bi(111) BL at $B=1$ eV as a function of the SOC strength. Numbers below each panel indicate the percentage of the 
strength of SOC in unit of $\xi_{\rm Bi}$. Black (red) lines stand for the bands with
$k_y=0$ ($k_x=0$). The numbers stand for the Chern numbers of the topmost valence bands.
\label{fig:berry}
}
\end{figure*}

The peculiarity of the topologically nontrivial phases with nonzero $\mathcal{C}_\text{s}$ lies in their finite SHC.
It can be understood in an intuitive way following two equations below:
\begin{equation}
\begin{split}
{\rm AHC}&=\sigma^\uparrow+\sigma^\downarrow \\
{\rm SHC}&=\sigma^\uparrow-\sigma^\downarrow 
\end{split}
\end{equation}
where $\sigma^\uparrow$ ($\sigma^\downarrow$) denotes the conductivity of the majority (minority) electrons
in units of $e^2/h$ ($e/4\pi$) for the AHC (SHC), respectively.
These relations can be defined and hold true only when the spin-flip band transitions due to spin-non-conserving part of SOC are absent. 
In the trivial insulator phase, both the AHC and SHC are zero since $\mathcal{C}_+=\mathcal{C}_-=0$. 
When time-reversal symmetry is not broken,~i.e.~for the Bi BL in the TI phase, it implies that 
$\sigma^\uparrow=-\sigma^\downarrow=-1$,
leading to zero AHC and to a quantized  SHC of $-2 e/4\pi$.\cite{Murakami:2006} 
Nevertheless, due to the spin-flip part of SOC, the SHC of the  Bi BL in the TI phase is reduced to 
about $0.7\ e/4\pi$, as we found in our previous work (Fig.~3 in Ref.~[\onlinecite{Zhang:2012}]).
In the QAH phase with $\mathcal{C}=-2$, $\mathcal{C}_+=\mathcal{C}_-=-1$, and, correspondingly, 
$\sigma^\uparrow=\sigma^\downarrow$, leading to zero SHC without the spin-flip transitions. However, 
the exchange splitting in the system leads to the fact that the spin-flip scattering between majority and minority 
states is not balanced, leading to a small but finite SHC, as shown in Fig.~\ref{fig:fig2}(b) for Bi BL at 
$B=0.5$ eV.

On the contrary, for the QAH phase with the Chern number $\mathcal{C}=+1$, the majority spin channel is switched off since 
$\mathcal{C}_+=0$, 
and the resulting SHC is enhanced in magnitude as compared to the QAH phase with $\mathcal{C}=-2$, see Fig.~\ref{fig:fig2}(b). As we
artificially suppress the spin-flip band transitions by switching off the spin-non-conserving part of SOC in our calculations, 
we observe that the value of the SHC acquires a quantized value of the magnitude of $-\frac{e}{4\pi}$ (not shown), despite the fact that the area of the $\mathcal{C}=+1$ phase shrinks as the electronic structure of the bilayer is modified when the spin-flip SOC is switched off.
That is, we demonstrated that it is possible to achieve a coexistence of the QAHE and quantized (in the spin-conserving  sense) spin Hall effect by applying a strong exchange field (at constant SOC strength) to a topological insulator, trivial insulator, 
or even a metal at $B=0$. We call such an emergent topological phase a {\it quantum spin Chern insulator} phase.

The variation of the Chern numbers in the phase diagram of the Bi BL can be explained by the Chern number exchange mechanism at 
the critical points, as discussed in the previous section.~\cite{Bellissard:1995} For phase transitions at constant $\xi$ and varying $B$  the Chern numbers 
are changed by $+1$ (for intermediate $\xi$) and $-2$ (for large $\xi$), Fig.~\ref{fig:fig2}(a), corresponding to 
linear and quadratic dispersions at the critical points, respectively, as confirmed by our calculations (not shown).
More interestingly, the Chern number is changed by $-3$ as we go from the $\mathcal{C}=+1$ phase to the $\mathcal{C}=-2$ 
phase at a constant large $B$ when varying the strength of SOC.
Our analysis reveals that this transition can be decomposed into three steps, at each of which the topology of the Berry curvature distribution
in the Brillouin zone changes, as analyzed in the following, Fig.~\ref{fig:berry} ($B=1$ eV).

We focus on the neighborhood of the $\Gamma$-point, since most of the contribution to the variation of the Berry curvature and
the Chern number exchange comes from the coupling of the highest occupied bands to the lowest unoccupied bands in this region.
For $\xi\le0.79\,\xi_{\rm Bi}$ in the $\mathcal{C}=+1$ phase the distribution of Berry curvature exhibits a  hot-loop structure with two pronounced singularity-like 
points at the $k_x=0$ axis. Enhancing the SOC strength brings down the conduction band and results in a singularity directly at $\Gamma$-point as the bands touch each other. Upon reopening the gap at the touching $\Gamma$-point the
Chern number of the valence band is changed from $+1$ to $+2$ ($\xi=0.80\,\xi_{\rm Bi}$ in Fig.~4). If the valence band at this $\xi$
was separated by a global gap from the conduction band, the Chern number of the system would be $+2$.
As the $\xi$ is increased further, the conduction band goes 
further down in energy, while the point of band crossing with the valence band and corresponding singularities in the Berry curvature split and 
move away from the $\Gamma$-point, resulting in a hot loop and four ``monopoles" at $\xi=0.82\,\xi_{\rm Bi}$.  
The Chern number of the valence band at each of such singular points is changed by $-1$, therefore, the resulting QAH phase for 
 $\xi\geq 82\%$ of $\xi_{\rm Bi}$ has the Chern number $\mathcal{C}=-2$. At 82\% of the Bi SOC strength the band structure at the
 $\Gamma$-point is almost indistinguishable from that at 79\%, see Fig.~4, although the Chern number, sign of the Berry curvature,
 and orbital character of valence and conduction bands have changed completely.

\subsection{Tuning $\xi$ in Bi bilayers: alloying with Sb}

We turn now to the question of how to achieve tuning of SOC strength and inducing a finite exchange field in a Bi(111) bilayer.
Tuning the strength of SOC can be achieved by alloying Bi with its isoelectronic but lighter element Sb, 
as demonstrated in~e.g.~BiTl(S$_{1-\delta}$Se$_\delta$)$_2$ where Se was substituted with the isoelectronic S.~\cite{Xu:2011}
In Fig.~\ref{fig:fig2}(c) we show the calculated phase diagram of Sb(111) BL with respect to the magnitude
of the exchange field $B$ and the strength of SOC $\xi$.   
The general features of the phase diagram of Sb BL are similar to those of the Bi BL. For instance, at weak SOC strength
and small magnetic field, both bilayers are trivial insulators, while QAH phases emerge with increasing $\xi$ 
and $B$. The QAH phase with Chern number $\mathcal{C}=-2$ emerges in Sb BL for $B \geq 0.4$~eV if the SOC strength 
of Sb is scaled by about three times of its atomic value, to reach $\xi\approx 2.7$~eV, which is very close to $\xi_{\rm Bi}$.

Clearly, there exist also significant differences between the phase diagrams in Fig.~3(a) and (c). 
First, the TI phase with time-reversal symmetry at $B=0$ is reached in Sb BL only when the SOC strength is scaled above
305\% of its atomic value, $\xi\geq 2.75$~eV, which is larger than the critical value in Bi BLs $\xi_{\rm Bi}\approx 1.9$ eV. 
Second, the boundary between the trivial insulator and the metallic phases is moved towards larger $\xi$ and $B$ regime in
Sb BL, as compared to Bi BL. 
Third, and most important, in the Sb BL in the intermediate $\xi$ regime ($0.8\,\xi_{\rm Sb}\leq \xi \leq 1.8\,\xi_{\rm Sb}$) and in a  larger exchange field, the QAH phase bares
a Chern number with the sign opposite to that in the Bi BL: $\mathcal{C}=-1$ for Sb as compared to $\mathcal{C}=+1$ for Bi.
We observe that the occurrence of the $\mathcal{C}=-1$ QAH phase is extremely sensitive to the fine details of the electronic structure
of the bilayer. For instance, at $B=1$ eV, a ``$-1$" QAHE phase is also present in a small region of the SOC strength
$2.6\,\xi_{\rm Sb}\leq \xi \leq 2.65\,\xi_{\rm Sb}$. Moreover, for $1.80 \,\xi_{\rm Sb}\leq \xi \leq 2.60\,\xi_{\rm Sb}$ and 
$B\approx 1$~eV, the Sb bilayer is gapped with a tiny gap of a few meV, which can not be measured in a realistic experiment 
in which disorder and temperature effects are inevitable. 

The reason for the differences in the phase diagrams of Sb(111) and Bi(111) BLs can be attributed to the 
difference in the fine details of their electronic structure. 
For instance, the magnitude of the $V_{pp\sigma}$ nearest neighbor hopping parameter
in Bi accounts to 80\% of that in Sb, while the values of the $V_{pp\sigma}$ next nearest neighbor hopping parameter in Bi is only half of 
that in Sb, as listed in Ref.~[\onlinecite{Liu:1995}].
This underlines that, to a certain extent, Bi$_{1-x}$Sb$_x$ alloys cannot be treated within a simple SOC strength scaling picture,  
and we suspect that more interesting and non-trivial topological phases might occur in these alloys in a strong 
exchange field due to the competition between $\mathcal{C}=\pm 1$ phases, which have to be captured from more accurate
first principles calculations.

The QAH phase with $\mathcal{C}$=$-1$ found in Sb BL (Fig.~\ref{fig:fig2}(c)) is topologically different from 
the QAH phase with $\mathcal{C}$=$+1$ in Bi BL (Fig.~\ref{fig:fig2}(a-b)).
According to the bulk-edge correspondence, the bulk properties (Chern numbers) are reflected in
the chirality of the edge states in a finite system.
Obviously, the chiralities of the edge states for the $\mathcal{C}=\pm1$ QAH phases are opposite to each other, as shown in 
the insets of Fig.~\ref{fig:fig2}(a) and Fig.~\ref{fig:fig2}(c).
That is, the edge state located on the upper edge of a Sb(111) ribbon (inset of Fig.~\ref{fig:fig2}(c)) is right-propagating,
while that in a Bi(111) ribbon (inset of Fig.~\ref{fig:fig2}(a)) is left-propagating.  
The SHC is of finite magnitude for both $\mathcal{C}=\pm 1$ phases (Fig.~\ref{fig:fig2}(b) for $\mathcal{C}=+1$), but of opposite 
sign due to different chiralities of the edge states. Overall, the $\pm 1$ QAHE phases in Bi and Sb BLs are the {\it quantum spin
Chern insulator} phases with opposite values of the anomalous and spin Hall conductivities.

\subsection{Tuning B in Bi bilayers: magnetic substrates and doping}

In order to induce a finite exchange field in Bi/Sb(111) bilayers, two realistic ways can be suggested: (i) either by depositing the 
bilayers on top of a suitable magnetic insulating substrate whose surface layer exhibits an in-plane ferromagnetic spin structure,
or (ii) by doping the systems with magnetic atoms. We first briefly consider (i). 

Europium chalcogenides (EuX, X=O, S, Se, Te) are magnetic semiconductors with diverse magnetic ordering and a wide range of 
lattice constants.~\cite{Souza-Neto:2009}
We report here on first principles calculations of Bi(111) bilayer on top of EuSe slabs along (111) direction 
terminated with Se atoms, Fig.~\ref{fig:fig2}(d), assuming ferromagnetic order in EuSe with magnetization direction perpendicular
to the slab, but the conclusions below also hold for other europium chalcogenides. 
EuSe has cubic structure with the lattice constant of $6.19$~\AA, which provides the smallest lattice mismatch of 
around 3\% to Bi(111) BL as compared to other europium chalcogenides. To describe Eu $4f$-electrons properly, we employed the
LDA+$U$ scheme with $U=7.0$~eV and $J=1.2$ eV.
As shown in Fig.~\ref{fig:fig2}(d), the states of the system around $E_F$ are originated from Bi bands, with the largest exchange 
splitting induced by hybridization with the magnetic substrate accounting to about 60 meV. Such magnitude of the exchange splitting
is definitely not enough to reach the QAH phase in Fig.~\ref{fig:fig2}(a), as verified by explicit AHC calculations (not shown), although it might be
used to probe the time-reversal broken TI phase in the system, if the Fermi energy of the hybrid systems
can be tuned to locate in the gap.
The small exchange splitting in the BL is due to the hybridization with the half-filled shell of $4f$  Eu electrons, which is quite
localized leading to insignificant exchange interactions in the $6p$ orbitals of Bi atoms. 
In this sense, using
a substrate with partially filled $d$-shell,~e.g.~MnSe,\cite{Luo:2012} might be favorable. 

On the other hand, doping with or adsorption of $3d$, $4d$ or $5d$ transition-metal atoms can induce strong exchange splitting and 
hence a non-trivial QAH effect in TIs, as shown in Refs.~[\onlinecite{Liu:2008, Yu:2010, Qiao:2010, Zhang:2012a}]. To demonstrate that
this is indeed the case for Bi bilayer, here we consider the case of Fe adatoms.
Compared to graphene, Bi(111) bilayer has a larger in-plane lattice constant ($4.52$~\AA, as compared to $2.46$~\AA~in~graphene). 
Our structural relaxations show that the Fe atoms can be stabilized at the hollow site of the hexagonal
plaquette in the middle of two Bi atomic layers at the spatial inversion point of the staggered honeycomb lattice.
Fig.~\ref{fig:fig2}(e) shows the electronic structure of Fe adatoms in p($2\times 2$) superstructure on Bi(111) BL, together with the calculated
anomalous Hall conductivity. Among two global gaps which are formed upon Fe deposition, the band gap at the Fermi energy is trivial, 
while the band gap at about 0.6 eV below $E_F$ exhibits a non-zero Chern number $\mathcal{C}=+2$ (Fig.~\ref{fig:fig2}(e)).
This is a new QAH phase, as compared to the phase diagram of Bi BL in Fig.~\ref{fig:fig2}(a), whose emergence
is due to a strong hybridization between the Fe $d$- and Bi $p$-states in the vicinity of the Fermi energy, in analogy to the case of 
graphene.\cite{Zhang:2012a, Hu:2012}
A similar mechanism also plays out for the p($2\times 2$) Co doped Bi(111) BLs, in which case the QAH gap is opened at about 0.3~eV below $E_F$
and the Chern number is $+2$.
Furthermore, varying the strength of SOC on Bi atomic sites can drive the system into yet different topological phases.
For instance, as exemplified in Fig.~\ref{fig:fig2}(f), if we scale the strength of SOC of Bi atoms to 20\% of their atomic value, 
the gap at $E_F$ of Fe doped Bi BLs acquires non-trivial topological properties and displays a $\mathcal{C}=-2$ QAH phase. 
This manifests that the system of $3d$ transition-metal adatoms/dopants on BiSb alloyed films definitely presents a unique and rich
system in which exciting non-trivial phases could exist and could be  tunable, suggesting thus the necessity for further theoretical 
and experimental investigations.

\section{Summary}
In summary, using a generic tight-binding model constructed for buckled honeycomb lattice with $sp$-orbitals, 
we illustrated how to induce nontrivial topological phases by generating an effective complex
next nearest neighbor hopping by spin-orbit interaction. 
Various channels can be recognized for $p_z$-, $s$-, and \{$p_x,p_y$\}-bands, respectively.
We found that both spin and orbital degrees of freedom can be utilized to generate the complex hoppings, and verified
that for a Bi(111) bilayer the orbital angular momentum purification plays an important role in its nontrivial 
properties due to the strong spin-orbit coupling strength of Bi atoms. Moreover, we showed how the nontrivial QAH phases
can be induced by direct mixing of the two spin channels with corresponding exchange of the Chern number between
the bands of opposite spin.

Further, we demonstrated that quantum anomalous Hall effect in Bi(111) bilayer can be induced by a combined tuning 
of the strength of the spin-orbit interaction and the magnitude of an external exchange field. We observe that various topological phases
of the Bi(111) bilayer, arising when the time-reversal symmetry is broken can be characterized by Chern and spin Chern numbers, 
corresponding to experimentally measurable transverse charge and spin conductivities.
We showed that tuning the spin-orbit strength can be achieved by alloying Bi with Sb, with rich physics expected
in resulting Bi$_{1-x}$Sb$_x$ systems. In order to induce a finite exchange field in a Bi(111) bilayer, we examined the effect
of a magnetic substrate, used for deposition of the bilayer, and magnetic dopants. We observe that, a Bi(111) bilayer 
on top of insulating europium chalcogenides exhibits a small exchange splitting of the bands, resulting in a time-reversal broken topological
phase. On the other hand, we predict that a much larger splitting can be achieved upon introducing magnetic transition-metal 
impurities in
the system, with resulting sizeable QAH gaps in the spectrum of the composite system formed due to strong hybridization 
between the Bi $p$- and adatom's $d$-states. The position of these gaps and their topological properties can be further tuned
by tuning the spin-orbit strength of the Bi(111) bilayer via,~e.g.~alloying with Sb. 

\section{Acknowledgments}
We acknowledge insightful discussions with Timo Schena and Congjun Wu. 
This work was supported by the HGF-YIG Programmes VH-NG-513 and VH-NG-409, as well as 
by the DFG through Research Unit 912 and grant HE3292/7-1. Computational 
resources were provided by the J\"ulich Supercomputing Centre.


\begin{thebibliography}{99}
\bibitem{Liu:1995} Y. Liu, and R. E. Allen, Phys. Rev. B {\bf 52}, 1566 (1995). We neglected the parameters between the next
nearest neighbors in this paper, which are only relevant for the hoppings between bilayers.
\end{thebibliography}
\end{document}